\documentclass[12pt]{iopart}
\usepackage{graphicx}
\usepackage{iopams}
\usepackage{amssymb}
\usepackage{multirow}

\newcommand{\ped}[1]{\ensuremath{_{\rm #1}}}

\begin{document}
\title{THz electrodynamics of BaFe$_{1.91}$Ni$_{0.09}$As$_{2}$ film analyzed in the framework of multiband Eliashberg theory}
\author{G.A. Ummarino$^{1,2}$, A.V. Muratov$^3$, L.S. Kadyrov$^4$, B.P. Gorshunov$^4$, S. Richter$^{5,6}$, A. Anna Thomas$^{5,6}$, R. H\"uhne$^5$ and Yu.A. Aleshchenko$^3$}
\ead{aleshchenkoya@lebedev.ru}
\address{$^1$ Istituto di Ingegneria e Fisica dei Materiali, Dipartimento di Scienza Applicata e Tecnologia, Politecnico di
Torino, Corso Duca degli Abruzzi 24, 10129 Torino, Italy}
\address{$^2$National Research Nuclear University MEPhI (Moscow Engineering Physics Institute), Kashirskoe shosse 31, Moscow 115409, Russia}
\address{$^3$ P.N. Lebedev Physical Institute, Russian Academy of Sciences, Leninskiy Prospekt 53, Moscow 119991, Russia}
\address{$^4$ Moscow Institute of Physics and Technology, Institutskiy per. 9, Dolgoprudny, Moscow Region, 141700, Russia}
\address{$^5$ Institute for Metallic Materials, Leibniz IFW Dresden, Helmholtzstrasse 20, Dresden 01069, Germany}
\address{$^6$ School of Sciences, TU Dresden, 01062 Dresden, Germany}

\begin{abstract}
The temperature dependences of the plasma frequency, superfluid density and London penetration depth were determined from terahertz spectra of conductivity and dielectric permittivity of BaFe$_{1.91}$Ni$_{0.09}$As$_{2}$ film with critical temperature $T_c=19.6$~K. Part of experimental data were analyzed within a simple three-band Eliashberg model where the mechanism of superconducting coupling is mediated by antiferromagnetic spin fluctuations, whose characteristic energy $\Omega_{0}$ scales with $T_{c}$ according to the empirical law $\Omega_{0} = 4.65k_{B}T_{c}$, and with a total electron-boson coupling strength $\lambda_{tot} = 2.17$.
\end{abstract}
\pacs{74.70.Xa, 74.20.Fg, 74.25.Kc, 74.20.Mn}
\maketitle
\section{Introduction}
Iron based superconductors are nowadays among the most studied superconducting (SC) materials due to their potential technological impact and their intriguing fundamental multiband and pairing properties. The widely accepted pairing state of most iron based superconductors and, in particular, of the compounds based on doped BaFe$_2$As$_2$ (Ba122 family) is the fully gapped $s_\pm $ phase~\cite{Chubukov 2008,Mazin_spm}. However, a possibility of the broken time-reversal symmetry in the SC state has been shown recently~\cite{chiral} for instance in (Ba,K)Fe$_{2}$As$_{2}$ at high doping levels by means of muon spin-relaxation measurements. This indicates the presence of the chiral type order parameter of either $s+is$ or $s+id$ symmetry. Typically the Fermi surface in these materials at the optimal doping level consists of holelike pockets at the $\Gamma $ point and electronlike pockets centered at the $X(0,\pm\pi $) points of the Brillouin zone. The pairing state consists of an extended $s$-wave pairing with a sign reversal of the order parameter between different Fermi surface sheets and with interband coupling between hole and electron bands provided by antiferromagnetic spin fluctuations. Although this model is widely accepted, direct experimental confirmations are rare~\cite{Tors1}.

Many doped SC compounds of the Ba122 family, especially Co- and K-doped, can be grown as epitaxial thin films with decent quality. However optical studies of Ni-doped Ba122 epitaxial films have been reported only in~\cite{Yoon 2017}. Moreover, even for a bulk Ba(Fe,Ni)$_2$As$_2$ data on the SC gap are scarce~\cite {Chi 2009,Ding 2009,Gong 2010,Dressel 2011,Kuzmicheva 2016}.

In this paper, we apply the Eliashberg model to describe SC properties of the BaFe$_{1.91}$Ni$_{0.09}$As$_{2}$ film probed by the terahertz (THz) spectroscopy. Measurements in the THz region has proven to be of particular importance in detecting the presence of SC gaps in superconductors, including multiband systems~\cite {Dore 2013}.

\section{Experiment}
The nearly optimum electron doped BaFe$_{1.91}$Ni$_{0.09}$As$_{2}$ film with a thickness of 120$\pm 10$~nm was grown by pulsed laser deposition (PLD) on polished (001) CaF$_2$ substrate. More details on the sample preparation are in~\cite{Richter 2017,S Richter 2017,Shipulin 2018}. The results of measurement of the SC transition temperature $T_c$ of Ni-Ba122 film by an inductive technique~\cite{Gavrilkin 2014} are shown in Fig.~1. A sharp SC transition was found at $T_c=19.6$~K.

THz spectroscopic measurements of the Ni-Ba122 film on the CaF$_2$ substrate were carried out in the transmission mode with the Menlo pulsed time-domain THz spectrometer within the range of 10--50~cm$^{-1}$ (wavelengths 1~mm--200~$\mu $m) at different temperatures down to $T=5$~K with a home-made optical cryostat. The measured transmission coefficient and the phase shift of the electromagnetic radiation allow us to determine directly the spectra of real and imaginary parts of the complex dielectric permittivity $\hat\varepsilon (\omega )=\varepsilon _1(\omega )+i\varepsilon _2(\omega )$ and the complex optical conductivity $\hat\sigma (\omega )=\sigma _1(\omega )+i\sigma _2(\omega )$ without using the Kramers-Kronig relation~\cite{Pracht 2013}. The dielectric properties of the CaF$_2$ substrate were measured beforehand.

\section{Results}
In this work, we focus on the itinerant charge-carrier response contained within the THz spectral range. The spectra of dielectric permittivity and conductivity of BaFe$_{1.91}$Ni$_{0.09}$As$_{2}$ film both in normal (a,c) and SC states (b,d) are displayed in Fig.~2. The abrupt decrease in the low-frequency permittivity $\varepsilon _1(\omega ,T)$ at $T<T_c$ shown in Fig.~2(b) evidences the inductive response of the SC condensate. This behavior is fitted with the expression $\varepsilon _1\propto -(\omega _{p,s}/\omega )^2$ (solid line), where $\omega _{p,s}$ is the SC plasma frequency. This fit allows us to determine $\omega _{p,s}$. The solid lines in Fig. 2(a), (c) and (d) are only guides to eye. At high temperatures, the THz conductivity is nearly frequency independent (see Fig. 2(c)). At $T<T_c$, the THz conductivity decreases due to the opening of the SC gap Fig. 2(d). It is easily recognized that even at the lowest temperature (5 K), i.e. 0.25$T_c$, the gapping is not complete. According to~\cite{Lobo 2010}, the presence of a finite conductivity at low frequencies well into the SC state could be due to a gap anisotropy of the electron pocket~\cite{Chubukov 2009,Mishra 2009,Carbotte 2010,Muschler 2009}; impurity localized levels inside an isotropic gap~\cite{Shiba 1968,Rusinov 1969,Schachinger 1984}; or pair breaking due to interband impurity scattering in an $s_\pm $ symmetric gap~\cite{Vorontsov 2009,Nicol 1992}.

Figure 3 shows the calculated temperature dependences of the SC plasma frequency $\omega _{p,s}$ and the London penetration depth $\lambda _L$ (right scale) determined as $\lambda _L=c/\omega _{p,s}$~\cite{Basov 2005}. The temperature dependence of the superfluid density $n_s=[\lambda _{L}(0)/\lambda _{L}(T)]^2$ is shown in the inset. A rough estimate of $\lambda _{L}(0)=0.414$~$\mu $m can be made assuming a parabolic dependence for $\lambda _{L}(T)$ and using three low-temperature points of this dependence (at higher temperatures this parameter is affected by the sample geometry and fluctuations). Such behavior is compatible with nodes in the SC gap~\cite{Fischer 2010} or a multigap system ~\cite{Vorontsov 2009}. To describe these experimental findings, an Eliashberg theory for Ba(Fe,Ni)$_2$As$_{2}$ has been developed.

\section {The Model}
The electronic structure of the compound BaFe$_{1.91}$Ni$_{0.09}$As$_{2}$, being very similar to Co doped case, can be approximately described by a three-band model with one hole band (indicated in the following as band 1) and two electron bands (indicated in the following as bands 2 and 3)~\cite{Tors2}. In this way the gap of the hole band, $\Delta_{1}$, has opposite sign to the two gaps residing on the electron bands, $\Delta_{2}$ and $\Delta_{3}$. The interband coupling between hole and electron bands ($s_\pm $ wave model \cite{Aigor,Mazin_spm}) is mainly provided by antiferromagnetic spin fluctuations (\textit{sf}), while phonons can be responsible for the intraband coupling (\textit{ph})~\cite{Mazin_spm}. The antiferromagnetic spin fluctuation coupling between bands with the same type of charge carriers (holes with holes and electrons with electrons) is zero while the phonon coupling is negligible \cite{Aigor,Mazin_spm}.
We assume that the symmetry of this system is $s_\pm $ for experimental and theoretical reasons. The experimental reasons are that almost all iron pnictides are $s_\pm $ and Ba(Fe,Ni)$_2$As$_{2}$ is very similar to Co doped case that is thought to be $s_\pm $~\cite{ummaco}. From a theoretical point of view, we use the theory developed in~\cite{ummaco1} where, within a Ginzburg-Landau formalism, the necessary and sufficient conditions to realize a $s_\pm $ state ($\pi$ state) are established analytically. It is necessary to premise that in our model, as we will see, the electron-boson coupling is just from antiferromagnetic spin fluctuations because we neglect the phonons. We will find the coupling constants $\lambda_{12}=-0.91$ and $\lambda_{13}=-1.356$ (they are positive in the paper because we, as usually, change the sign inside the equations) and $\lambda_{23}=0$. Of course $\lambda_{23}$ differs from zero and is positive but small because is connected with the phonon contribution. We have verified in the past~\cite{Umma1} that small phononic couplings (very small compared to the coupling value of antiferromagnetic spin fluctuations) do not change the result of the numerical simulations and, for simplicity and to reduce the number of free parameters, they have been set to zero. In~\cite{ummaco1} the functions $G_{1}$, $G_{2}$ and $G_{3}$ are definied as $G_{1}=\frac{\lambda_{12}}{\mid \lambda_{23} \mid} \frac{\mid\psi_{1}\mid}{\mid\psi_{2}\mid}$, $G_{2}=sign(\lambda_{23})$ and $G_{3}=\frac{\lambda_{13}}{\mid \lambda_{23} \mid} \frac{\mid\psi_{3}\mid}{\mid\psi_{2}\mid}$. $G_{1}$ and $G_{3}$ are negative and large and $G_{2}$ is positive and small since $\lambda_{23}\approx 0$ and the $\psi_{i}$ are more or less of the same order of magnitude because they are proportional to square root of superfluid densities $n_{i}$ in the bands 1, 2 and 3 (that we have calculated for obtaining the penetration depth). We are in the low left corner of Fig.~2 of the same paper where the solution $\pi $ is stable.
To calculate the gaps and the critical temperature within the $s_\pm $ wave three-band Eliashberg equations~\cite{Eliashberg,Chubukov,Tors3}, one has to solve 6 coupled equations for the gaps $\Delta_{i}(i\omega_{n})$ and the renormalization functions $Z_{i}(i\omega_{n})$, where $i$ is a band index (that ranges between $1$ and $3$) and $\omega_{n}$ are the Matsubara frequencies. The imaginary-axis equations \cite{Umma1,Umma2,Umma3} read:
\begin{eqnarray}
&&\omega_{n}Z_{i}(i\omega_{n})=\omega_{n}+ \pi T\sum_{m,j}\Lambda^{Z}_{ij}(i\omega_{n},i\omega_{m})N^{Z}_{j}(i\omega_{m})+\nonumber\\
&&+\sum_{j}\big[\Gamma^{N}\ped{ij}+\Gamma^{M}\ped{ij}\big]N^{Z}_{j}(i\omega_{n})
\label{eq:EE1}
\end{eqnarray}
\begin{eqnarray}
&&Z_{i}(i\omega_{n})\Delta_{i}(i\omega_{n})=\pi
T\sum_{m,j}\big[\Lambda^{\Delta}_{ij}(i\omega_{n},i\omega_{m})-\mu^{*}_{ij}(\omega_{c})\big]\times\nonumber\\
&&\times\Theta(\omega_{c}-|\omega_{m}|)N^{\Delta}_{j}(i\omega_{m})
+\sum_{j}[\Gamma^{N}\ped{ij}-\Gamma^{M}\ped{ij}]N^{\Delta}_{j}(i\omega_{n})\phantom{aaaaaa}
 \label{eq:EE2}
\end{eqnarray}
where $\Gamma^{N}\ped{ij}$ and $\Gamma^{M}\ped{ij}$ are the scattering rates from non-magnetic and magnetic impurities, $\Lambda^{Z}_{ij}(i\omega_{n},i\omega_{m})=\Lambda^{ph}_{ij}(i\omega_{n},i\omega_{m})+\Lambda^{sf}_{ij}(i\omega_{n},i\omega_{m})$ and
$\Lambda^{\Delta}_{ij}(i\omega_{n},i\omega_{m})=\Lambda^{ph}_{ij}(i\omega_{n},i\omega_{m})-\Lambda^{sf}_{ij}(i\omega_{n},i\omega_{m})$
where
\[\Lambda^{ph,sf}_{ij}(i\omega_{n},i\omega_{m})=2
\int_{0}^{+\infty}d\Omega \Omega
\alpha^{2}_{ij}F^{ph,sf}(\Omega)/[(\omega_{n}-\omega_{m})^{2}+\Omega^{2}]. \]
$\Theta$ is the Heaviside function and $\omega_{c}$ is a cutoff
energy.  The quantities $\mu^{*}_{ij}(\omega\ped{c})$ are the elements of the $3\times 3$
Coulomb pseudopotential matrix. Finally,
$N^{\Delta}_{j}(i\omega_{m})=\Delta_{j}(i\omega_{m})/
{\sqrt{\omega^{2}_{m}+\Delta^{2}_{j}(i\omega_{m})}}$ and
$N^{Z}_{j}(i\omega_{m})=\omega_{m}/{\sqrt{\omega^{2}_{m}+\Delta^{2}_{j}(i\omega_{m})}}$.
The electron-boson coupling constants are defined as
$\lambda^{ph,sf}_{ij}=2\int_{0}^{+\infty}d\Omega\frac{\alpha^{2}_{ij}F^{ph,sf}(\Omega)}{\Omega}$.
The solution of equations \ref{eq:EE1} and \ref{eq:EE2} requires a huge number of input parameters (18 functions and 27 constants), i.e.:
i) nine electron-phonon spectral functions $\alpha^{2}_{ij}F^{ph}(\Omega)$; ii) nine electron-antiferromagnetic spin fluctuation spectral functions, $\alpha^{2}_{ij}F^{sf}(\Omega)$; iii) nine elements of the Coulomb pseudopotential matrix $\mu_{ij}^{*}(\omega_{c})$;
iv) nine nonmagnetic $\Gamma^{N}\ped{ij}$ and nine paramagnetic $\Gamma^{M}\ped{ij}=0$ impurity-scattering rates.
However, some of these parameters can be extracted from experiments and some can be fixed by suitable approximations.
In particular, we refer to experimental data taken on high quality films, so we can rather safely assume a negligible disorder: the scattering from non-magnetic impurities $\Gamma^{N}\ped{ij}$ can thus be taken to be zero. The same can be done for the scattering rate from magnetic impurities: $\Gamma^{M}\ped{ij}=0$.
At least as a starting point, let us make further assumptions that have been shown to be valid for iron pnictides \cite{Umma1,Umma2, Umma3}. Following ref. \cite{Mazin_spm}, we can thus assume that: i) the total electron-phonon coupling constant is small (the upper limit of the phonon coupling in the usual iron-arsenide compounds is  $\approx0.35$ \cite{Boeri2}); ii) phonons mainly provide \emph{intra}band coupling so that $\lambda^{ph}_{ij}\approx0$; iii) spin fluctuations mainly provide \emph{interband coupling between hole and electron bands}, so that $\lambda^{sf}_{ii}\approx0$. Moreover, we put in first approximation the phonon contribution to \emph{intra}band coupling equal to zero so that $\lambda^{ph}_{ii}=0$ so as, following Mazin \cite{Aigor}, the Coulomb pseudopotential matrix: $\mu^{*}_{ii}(\omega\ped{c})=\mu^{*}_{ij}(\omega\ped{c})=0$~\cite{Aigor,Umma1,Umma2,Umma3}. As we discussed before the maximum value of the total electron-phonon coupling is estimated to be under $0.35$ more or less similar to contribution, with opposite sign of Coulomb pseudopotential, so \textit{in the first approximation}, and for reducing the number of free parameters, we put the two contributions equal to zero because they cancel each other out. Of course this doesn't means that the phonons are absent but just that the final result in the calculus of a lot of physical properties is not influenced by the their presence. Within these approximations, the electron-boson coupling-constant matrix $\lambda_{ij}$ becomes:
\cite{Aigor,Umma1,Umma2,Umma3}:
\begin{equation}
\vspace{2mm} %
\lambda_{ij}= \left (
\begin{array}{ccc}
  0                 &         \lambda^{sf}_{12}                  &               \lambda^{sf}_{13}            \\
  0                &               \lambda^{sf}_{21}=\lambda^{sf}_{12}\nu_{12}               &                0            \\
  \lambda^{sf}_{31}=\lambda^{sf}_{13}\nu_{13} &  0  & 0 \\
\end{array}
\right ) \label{eq:matrix}
\end{equation}
where $\nu_{ij}=N_{i}(0)/N_{j}(0)$, and $N_{i}(0)$ is the normal
density of states at the Fermi level for the $i$-th band.
The coupling constants $\lambda_{ij}^{sf}$ are defined through the electron-antiferromagnetic spin fluctuation spectral functions (Eliashberg functions) $\alpha^2_{ij}F_{ij}^{sf}(\Omega)$. Following refs. \cite{Umma1,Umma2,Umma3} we choose these functions to have a Lorentzian shape, i.e.:
\begin{equation}
\alpha_{ij}^2F^{sf}_{ij}(\Omega)= C_{ij}\big\{L(\Omega+\Omega_{ij},Y_{ij})-
L(\Omega-\Omega_{ij},Y_{ij})\big\},
\end{equation}
where
\[
L(\Omega\pm\Omega_{ij},Y_{ij})=\frac{1}{(\Omega \pm\Omega_{ij})^2+Y_{ij}^2}
\]
and $C_{ij}$ are normalization constants, necessary to obtain the proper values of $\lambda_{ij}$, while $\Omega_{ij}$ and $Y_{ij}$ are the peak energies and the half-widths of the Lorentzian functions, respectively \cite{Umma3}.  In all the calculations we set $\Omega_{ij}=\Omega_{0}$, i.e. we assume that the characteristic energy of spin fluctuations is a single quantity for all the coupling channels, and  $Y_{ij}= \Omega_{0}/2$, based on the results of inelastic neutron scattering measurements \cite{Inosov}.

The peak energy of the Eliashberg functions, $\Omega_0$, can be directly associated to the experimental critical temperature, $T_c$, by using the empirical law $\Omega_{0}=2T_{c}/5$ that has been demonstrated to hold, at least approximately, for iron pnictides \cite{Paglione}. With all these approximations, necessary to reduce the number of free parameters, this is a more simple model that can still grasp the essential physics of iron compounds.
We use a cut-off energy $\omega_{c}=180$ meV and a maximum quasiparticle energy $\omega_{max}=200$ meV.

The factors $\nu_{ij}$ that enter the definition of $\lambda_{ij}$ (eq. 3) are unknown so we assume that they are equal to the Co doped \cite{Umma3} case so $\nu_{12}=1.12$ and $\nu_{13}=1$. In the same way we assume that also the coupling constant are similar to Co doped case \cite{Tors2} and we change lightly just a value for obtaining the correct critical temperature: $\lambda_{12}=0.910$, $\lambda_{13}=1.356$ and $\lambda_{23}=\lambda_{ii}=0$ for a total coupling constant $\lambda_{t}=2.17$.

We have calculated before, by solving Eliashberg equations on the imaginary axis for all temperatures under $T_{c}$, the temperature dependence of $\Delta_{i}(i\omega_{n=0})$ as is shown in Fig.~4 (dashed line) and then, using Pad\'{e} approximants, we calculate the values of the gaps $\Delta_{i}(T)$ just at low temperatures. At the end we solve the Eliashberg equations on the real axis (solid line) and we calculate the values of $\Delta_{i}(T)$ in all temperature range under $T_{c}$. Of course the values calculated by Pad\'{e} approximants~\cite{pade} and by real axis solution are the same. The difference between $\Delta_{i}(i\omega_{n=0})$ and $\Delta_{i}(T)$ born from the fact that, in the weak coupling regime (BCS limit: electron-boson coupling constant less than one), the imaginary part of the complex functions $\Delta_{i}(\omega,T)$ and $Z_{i}(\omega,T)$ can be neglected so the values are almost the same while in the strong coupling regime this is not possible~\cite{max}. The gap values at low temperatures are $\Delta_{1}=4.3$ meV, $\Delta_{2}=-2.8$ meV and $\Delta_{3}=-5.9$ meV.

\section{Calculation of the penetration depth}
 The penetration depth (or the superfluid density, see Fig.~5) can be computed starting from the gaps $\Delta_{i}(i\omega_{n})$ and the renormalization functions $Z_{i}(i\omega_{n})$ by
\begin{eqnarray}
\lambda^{-2}(T)=(\frac{\omega_{p}}{c})^{2} \sum_{i=1}^{3}w_{i}\pi T \sum_{n=-\infty}^{+\infty}\frac{\Delta_{i}^{2}(\omega_{n})Z_{i}^{2}(\omega_{n})}{[\omega^{2}_{n}Z_{i}^{2}(\omega_{n})+\Delta_{i}^{2}(\omega_{n})Z_{i}^{2}(\omega_{n})]^{3/2}}\label{eq.lambda}
\end{eqnarray}
where $w_{i}=\left(\omega_{p,i}/\omega_{p}\right)^{2}$ are the weights of the single bands, $\omega_{p,i}$ is the plasma frequency of the $i$-th band and $\omega_{p}$ is the total plasma frequency. Here, we can only act on the weights $w^\lambda_i$ in order to adapt the calculation to the experimental $\lambda_L(T)$ \cite{Tors4}. The multiplicative factor that involves the plasma frequencies derives from the fact that the low-temperature value of the penetration depth $\lambda_L(0)$ should, in principle, be related to the plasma frequency by $\omega_p=c/\lambda_L(0)$. This is strictly valid only for a clean uniform superconductor at $T=0$ if strong-coupling effects (or, more generally, Fermi-liquid effects) are negligible. In our case $w_{1}=0.05$, $w_{2}=0.80$ and $w_{3}=0.15$. We find from the fit that $\lambda(0)=0.414$ $\mu m$.\\

\section{Discussion}
The three-band Eliashberg model adequately describes experimental temperature dependences of the superfluid density and penetration depth of the BaFe$_{1.91}$Ni$_{0.09}$As$_{2}$ film. The calculated values of the larger gaps fall within the range reported in~\cite{Chi 2009,Ding 2009,Gong 2010,Dressel 2011,Kuzmicheva 2016}, the smaller gap correlates well with that found in~\cite{Kuzmicheva 2016}. The London penetration depth found in~\cite{Yoon 2017} for a BaFe$_{1.9}$Ni$_{0.1}$As$_{2}$ film was somewhat smaller ($\lambda _{L}\sim 226\pm 20$~nm) compared to the present study, which can be ascribed to the different composition of our film.
We can check the  free input parameters values if we try to reproduce also the upper critical field $B_{c2}$ in relation to temperature. In literature there are experimental data~\cite{Ni} relative to a bulk single crystal sample with a critical temperature between $18.8$ K and $19.4$ K and therefore very close to that of our film. There are also experimental data relative to epitaxial thin films~\cite{S Richter 2017} with $T_{c}=17.2$ K. From the particular shape of the experimental curve that represents the temperature dependence of the upper critical field it can be understood that the system must be multiband~\cite{Umma2}.
The multiband Eliashberg model developed above can also be used to explain the experimental results of upper critical field measurements \cite{Umma2,Sud,Bc2} as a function of temperature. For the sake of completeness, we give here the linearized gap equations in the presence of a magnetic field, for a superconductor in the clean limit (negligible impurity scattering). In the following, $v_{Fj}$ is the Fermi velocity of band $j$, and $H_{c2}$ is the upper critical field:
\begin{eqnarray}
\omega_{n}Z_{i}(i\omega_{n})\hspace{-2mm}&=&\hspace{-2mm}\omega_{n}+\pi
T\sum_{m,j}\Lambda_{ij}(i\omega_{n}-i\omega_{m})\mathrm{sign}(\omega_{m})\nonumber\\
Z_{i}(i\omega_{n})\Delta_{i}(i\omega_{n})\hspace{-2mm}&=&\hspace{-2mm}\pi
T\sum_{m,j}[\Lambda_{ij}(i\omega_{n}-i\omega_{m})-\mu^{*}_{ij}(\omega_{c})]\cdot \nonumber\\
& &
\hspace{-2mm}\cdot\theta(|\omega_{c}|-\omega_{m})\chi_{j}(i\omega_{m})Z_{j}(i\omega_{m})\Delta_{j}(i\omega_{m})\nonumber
\end{eqnarray}
\begin{eqnarray}
\chi_{j}(i\omega_{m})\hspace{-2mm}&=&\hspace{-2mm}(2/\sqrt{\beta_{j}})\int^{+\infty}_{0}dq\exp(-q^{2})\cdot\nonumber\\
& & \hspace{-2mm}\cdot
\tan^{-1}[\frac{q\sqrt{\beta_{j}}}{|\omega_{m}Z_{j}(i\omega_{m})|+i\mu_{B}H_{c2}\mathrm{sign}(\omega_{m})}]\nonumber
\end{eqnarray}
with $\beta_{j}=\pi H_{c2} v_{Fj}^{2}/(2\Phi_{0})$.
In these equations the three bare Fermi velocities $v_{Fj}$ are the input parameters. The number of adjustable parameters can be reduced~\cite{Umma2} to one by assuming that, as in a free-electron gas, $v_{Fj}\propto N^{j}\ped{N}(0)$ so that $v_{F2}=\nu_{21} v_{F1}$ and $v_{F3}=\nu_{31} v_{F1}$. We find $v_{F1}=2.228\times 10^{5}$~m/s for single crystal and $1.963\times 10^{5}$~ m/s for epitaxial thin films. In Fig.~6 the results of the theoretical calculations are shown and we can see that, in both cases, the agreement with the experimental data is good.
\section{Conclusions}
Using THz spectroscopy, we measured the far-infrared optical conductivity of BaFe$_{1.91}$Ni$_{0.09}$As$_{2}$ film in the normal and SC states. We found a clear signature of the SC gap, but the conductivity does not vanish in the SC state. The three-band Eliashberg approach was applied to describe this feature as well as the temperature dependences of the superfluid density and London penetration depth. Different independent experimental data (upper critical magnetic field versus temperature) seems confirm the choice of input parameters in the Eliashberg equations. Our results provide a strong evidence that BaFe$_{1.91}$Ni$_{0.09}$As$_{2}$ is a multi-band superconductor with $s_\pm $ pairing mediated by antiferromagnetic spin fluctuations.
\section{ACKNOWLEDGMENTS}
G.A.U. acknowledges support from the MEPhI Academic Excellence Project (Contract No. 02.a03.21.0005). The work of A.V.M. and Yu.A.A. is carried out within the state assignment of the Ministry of Science and Higher Education of the Russian Federation (theme "Physics of high-temperature superconductors and novel quantum materials", No. 0023-2019-0005). LSK and BPG acknowledge support from the Ministry of Education and Science of the Russian Federation (Program 5 top 100). S.R., A.A.T. and R.H. acknowledge the financial support by the German Research Foundation (DFG) within the framework of the research training group GRK1621.\\

\begin{figure}[ht]
\begin{center}
\includegraphics[keepaspectratio, width=\columnwidth]{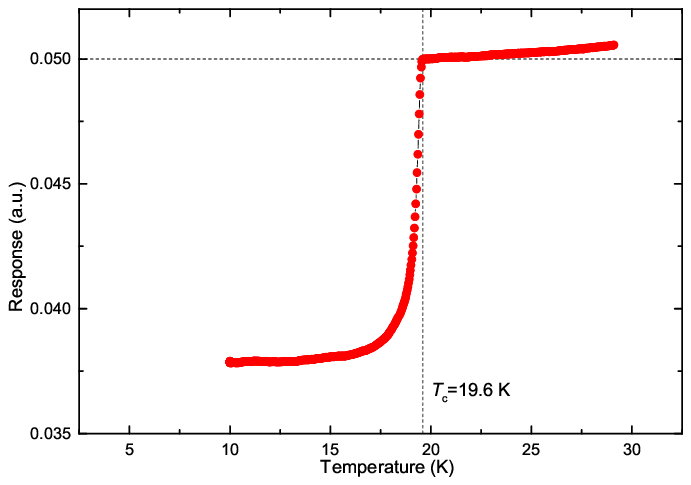}
\caption{(Color online) The temperature of SC transition $T_c$ of Ni-Ba122 film determined by the change in the magnetic susceptibility of the sample in inductive technique.}
\end{center}
\end{figure}

\newpage
\begin{figure}[ht]
\begin{center}
\includegraphics[keepaspectratio, width=\columnwidth]{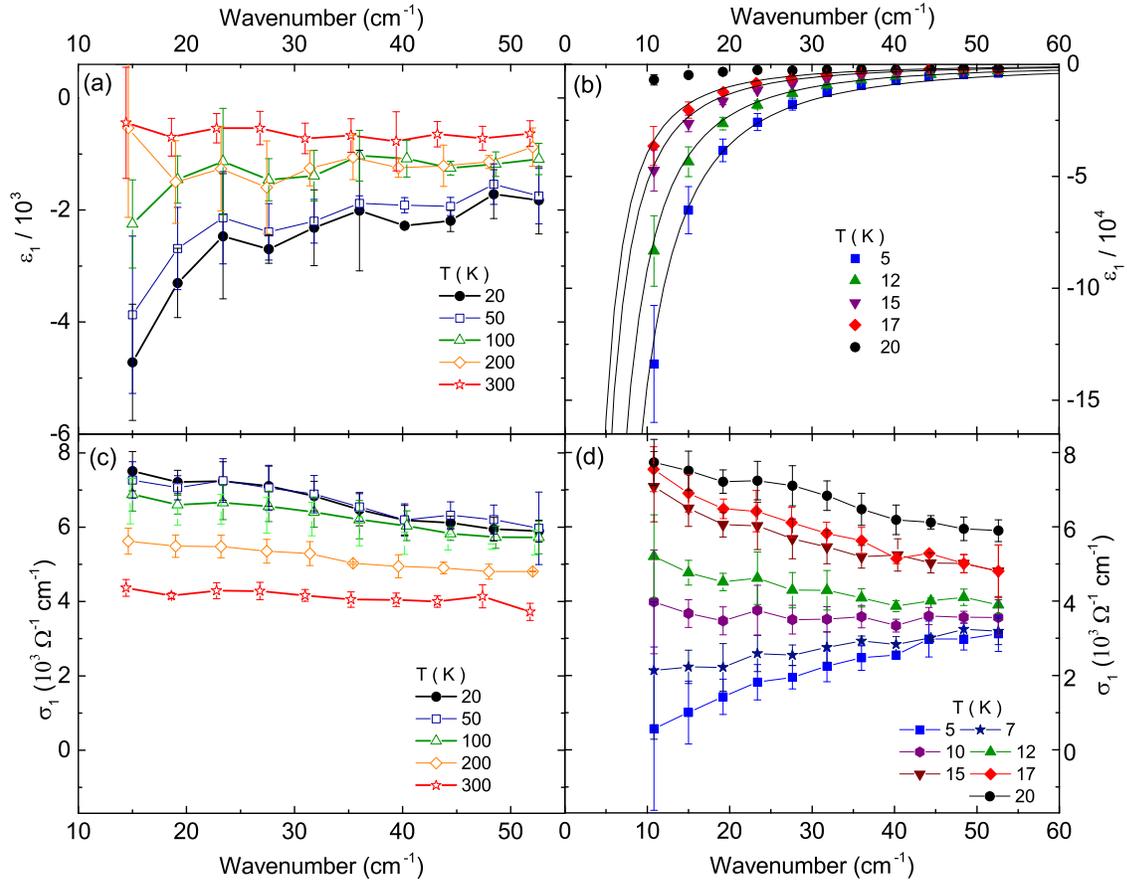}
\caption{(Color online) Spectra of the permittivity $\varepsilon _1$ and the conductivity $\sigma _1$ in the normal (a,c) and SC (b,d) phases. Solid lines in (b) show a fit of the spectra by the relation $\varepsilon _1\propto -(\omega _{p,s}/\omega )^2$.}
\end{center}
\end{figure}

\newpage
\begin{figure}[ht]
\begin{center}
\includegraphics[keepaspectratio, width=\columnwidth]{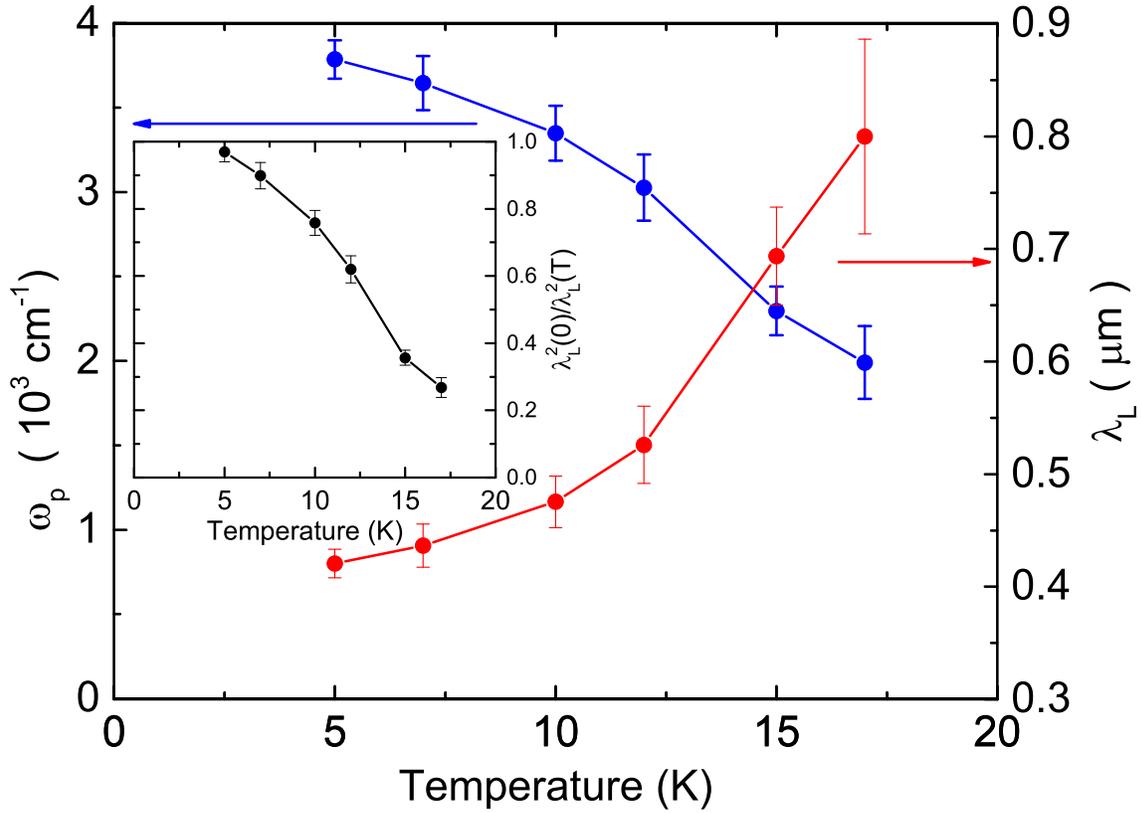}
\caption{(Color online) Temperature dependences of the plasma frequency of the SC condensate (left scale) and of the London penetration depth (right scale). The temperature dependence of superfluid density $n_{s}=[\lambda _{L}(0)/\lambda _{L}(T)]^2$ is shown in the inset.}
\end{center}
\end{figure}

\newpage
\begin{figure}[ht]
\begin{center}
\includegraphics[keepaspectratio, width=\columnwidth]{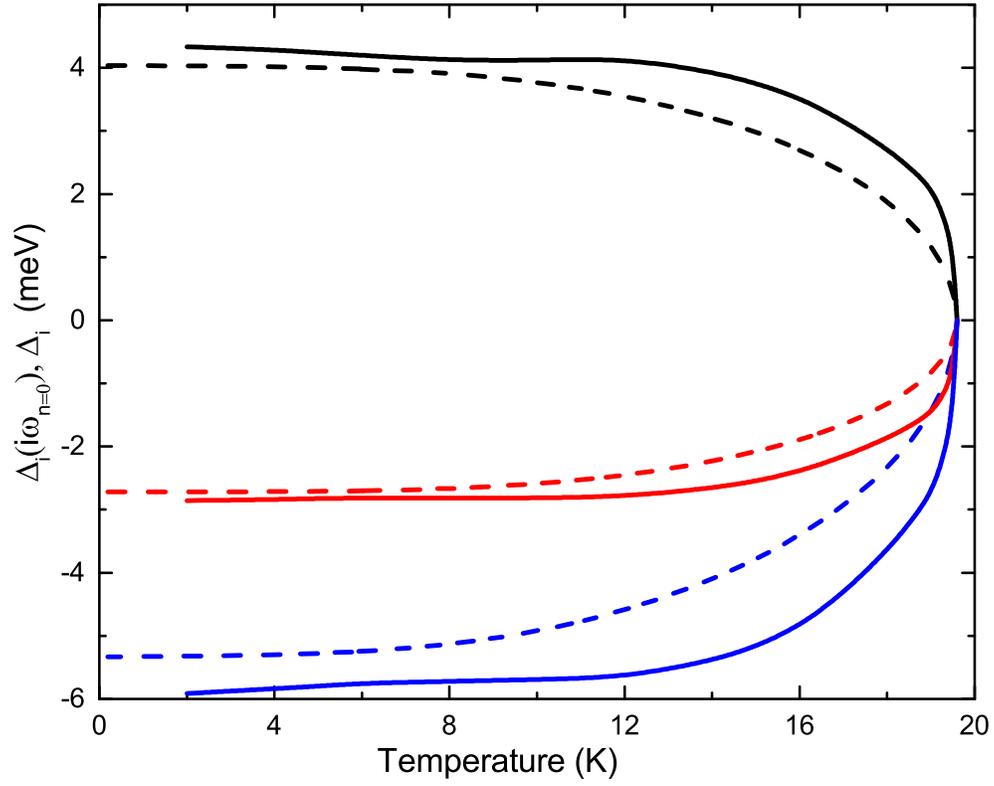}
\caption{(Color online)
The gaps $\Delta_{i}(i\omega_{n=0})$ as a function of temperature obtained by solving the Eliashberg equations on imaginary axis (dashed lines: black band 1, red band 2 and dark blue band 3) and real axis (solid lines).}
\end{center}
\end{figure}

\newpage
\begin{figure}[ht]
\begin{center}
\includegraphics[keepaspectratio, width=\columnwidth]{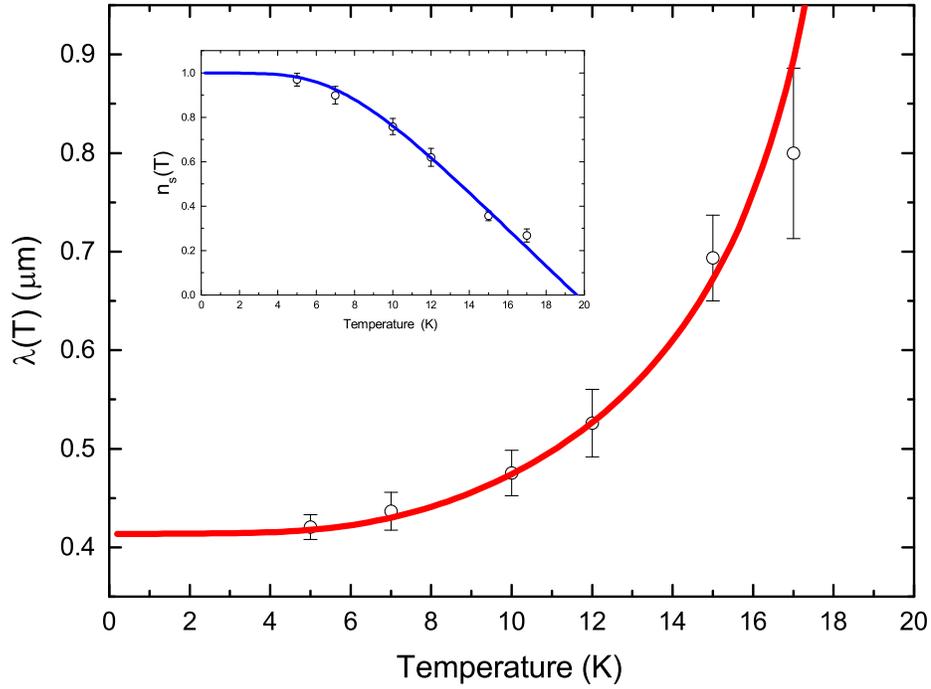}
\caption{(Color online)
The experimental penetration depth against the temperature and
the theoretical curve obtained by solving the Eliashberg equations.
In the inset the experimental superfluid density as a function of temperature is shown and
the theoretical curve obtained by solving the Eliahsberg equations.}
\end{center}
\end{figure}

\newpage
\begin{figure}[ht]
\begin{center}
\includegraphics[keepaspectratio, width=\columnwidth]{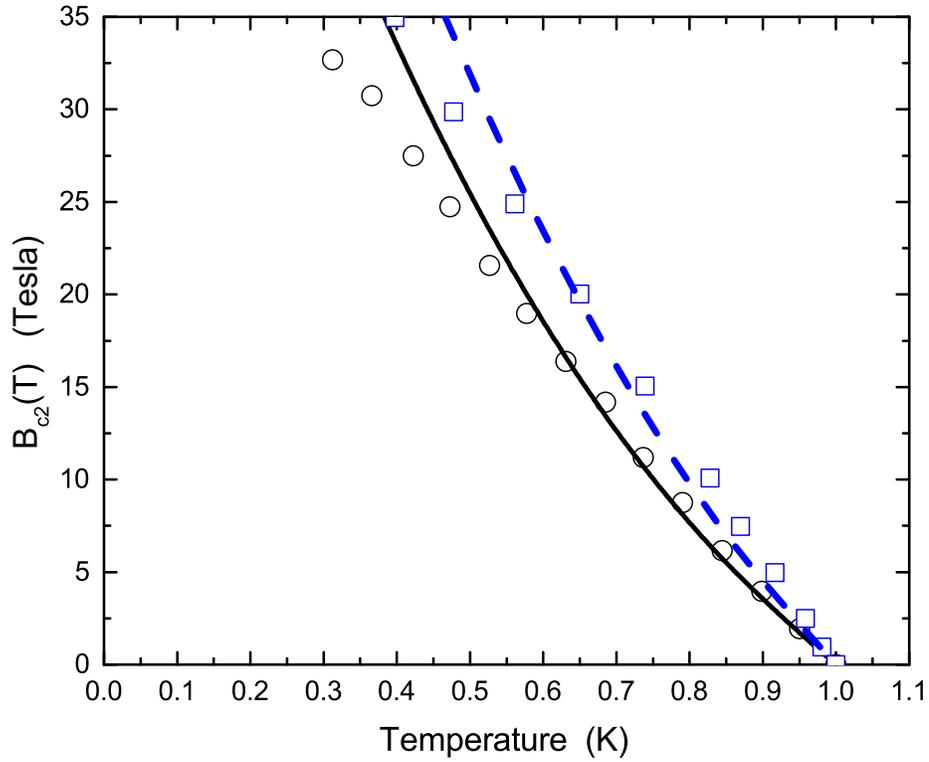}
\caption{(Color online)
The experimental upper critical field parallel to c axis versus temperature from \cite{Ni} (black open circles) and from \cite{S Richter 2017} (blue open squares). The theoretical curves obtained by solving the Eliahsberg equations are shown by black solid line and blue dashed line.
The experimental data are normalized to onset critical temperature of $18.8$ K and $17.6$ K.}
\end{center}
\end{figure}
\end{document}